\newcommand\vx{\vec{x}}
\newcommand\vr{\vec{r}}

\newcommand\vs{\vec{s}}
\newcommand\hk{\hat{k}}

\newcommand\hs{\hat{s}}
\newcommand\hz{\hat{z}}

\newcommand\hr{\hat{r}}

\newcommand\omegam{\Omega_{\rm m}}

\newcommand\oO{\mathcal{O}}
\newcommand\vz{\vec{z}}
\newcommand\eps{\epsilon}
\newcommand\muB{\mu_{\rm B}}
\newcommand\muM{\mu_{\rm M}}

\newcommand\Mpch{{\rm\; Mpc}/h}

\documentclass[useAMS,usenatbib]{mn2e}

\usepackage{lineno}

\usepackage{placeins}

\usepackage{graphicx} 
\usepackage{multirow}
\usepackage{array}

\usepackage{amsmath,amssymb}

\usepackage[T1]{fontenc}
\usepackage[latin9]{inputenc}
\usepackage{float}
\usepackage{graphicx}
\usepackage{esint}

\makeatletter
\DeclareFontEncoding{LGR}{}{}
\DeclareTextSymbol{\~}{LGR}{126}

\begin{document}

\title[Anisotropic 2PCF with FTs]{A new look at lines of sight: using Fourier methods for the wide-angle anisotropic 2-point correlation function}

\author{\makeatauthor}

\author[Slepian and Eisenstein]{Zachary Slepian\thanks{E-mail: zslepian@cfa.harvard.edu} and Daniel J. Eisenstein\thanks{E-mail: deisenstein@cfa.harvard.edu}\\
Harvard-Smithsonian Center for Astrophysics, Cambridge, MA 02138\\
}
\maketitle

\begin{abstract}
The anisotropic 2-point correlation function (2PCF) of galaxies measures pairwise clustering as a function of the pair separation's angle to the line of sight. The latter is often defined as either the angle bisector of the observer-galaxy-pair triangle or the vector from the observer to the separation midpoint. Here we show how to accelerate either of these
measurements with Fourier Transforms, using a slight generalization
of the Yamamoto et al. (2006) estimator in which each member of the pair
is used successively as the line of sight. We also present perturbation theory predictions for our generalized estimator including wide-angle corrections.
\end{abstract}

\section{Introduction}

Since the first redshift surveys in the late 1970s, the two-point
correlation function (2PCF) of galaxies has emerged as an important
probe of the cosmological parameters (e.g. Peebles 1980; Anderson et al. 2014). The 2PCF is often reported as a function solely of pair separation, $s$, corresponding to having been isotropically averaged around the observer. However, in reality the measured clustering depends both on the separation and on the line of sight to each galaxy in the pair. This breaking of isotropy, known as redshift space distortion (RSD), occurs because  galaxies are not comoving with
the background expansion of the Universe, but rather have peculiar velocities generated
both by virialization within clusters and by the growth of large-scale
structure. Consequently distance information along the line of sight, obtained by assuming all galaxies are comoving with the background expansion, is not fully accurate. The measured separation $s$ will therefore be faulty, more so the more the separation lies along the line of sight of either of the galaxies.

While RSD present a challenge for comparison with theory, they also offer an opportunity. If General Relativity (GR) is assumed, the anisotropic clustering probes the combination $f\sigma_8$, with $f$ the logarithmic derivative of the linear growth rate and $\sigma_8$ the clustering on $8\Mpch$ scales (Percival \& White 2009; Samushia, Percival \& Raccanelli 2012; Chuang \& Wang 2013; Oka et al. 2014). Alternatively, the velocities inferred from anisotropic clustering can be compared with those predicted by GR as a stringent test of the latter (Taruya et al. 2014; Taddei \& Amendola 2015).

As already noted, the observed clustering depends on both the separation and the angle between the separation and the line of sight to each galaxy in the pair. The full geometry is therefore a triangle formed by the observer and the pair of galaxies.  However in practice the geometry is often reduced from 3 parameters to 2 by defining an average line of sight. This does not lose much information if the typical triangle opening angle $\theta$ is small, as is commonly the case in present-day surveys. There are several approaches to defining an average line of sight. 

The crudest approach, known as flat-sky or plane-parallel, is using a single line of sight $\hz$ to the entire survey. This is appropriate if the angular scale of the survey is small. The Kaiser formula (1987) in Fourier space or the Hamilton formulae in real space (1992) were derived under this assumption using linear perturbation theory (PT) (see also Taylor \& Hamilton 1996; Nishioka \& Yamamoto 1999; Bharadwaj 2001).  These formulae predict the multipole moments of the clustering with respect to the cosine $\mu=\hz\cdot\hs$ of the angle between this single line of sight and the separation.  The parity of the Legendre polynomials demands that only even multipoles enter, and the expansion is relatively compact, involving only $l=0,2$ and $4$.  These formulae have been the dominant framework for interpreting anisotropic 2PCF or power spectrum measurements, often with some additional prescription for small-scale, non-linear effects such as fingers of God caused by collapse and virialization (Jackson 1972; Peacock \& Dodds 1996; Scoccimarro 2004; Taruya, Nishimichi \& Saito 2010). 

However, many surveys are now too wide in angle to make the flat-sky approximation.  A more accurate definition for the average line of sight is either the line of sight to a single pair member (SPM) (Yamamoto et al. 2006), the angle bisector of the triangle opening angle (Szalay, Matsubara \& Landy 1998; Matsubara 2000; Szapudi 2004; Yoo \& Seljak 2015), or the vector from the observer to the separation's midpoint (Yamamoto et al. 2006; Samushia, Branchini \& Percival 2015; Bianchi et al. 2015).  These methods will be the primary focus of this work.  They are all still approximate relative to tracking the full triangular geometry, but, as we will show, the information they lose is at $\oO(\theta^2)$ in the opening angle rather than $\oO(\theta)$ as in the flat-sky approach.  Indeed, it has been found empirically that for $\theta < 10^{\circ}$, there is very little difference between using the midpoint method and using the full triangular geometry (Samushia, Percival \& Raccanelli 2012; Beutler et al. 2012; Yoo \& Seljak 2015).  

It is well known that Fourier Transforms (FTs) can provide large
speed advantages for the computation of correlation functions.  It is
therefore desirable to be able to apply FTs to these more accurate methods
for the line of sight. However, applying FTs to compute these lines of sight is not transparent beause the line of sight varies throughout the
survey volume. FTs do have the disadvantage that they grid the data, losing some spatial information (see Slepian \& Eisenstein 2015, hereafter SE15, for further discussion). However, in contexts where number of catalogs or objects is more important than spatial precision,  FT techniques are an important complement to direct pair counting.

It has recently been shown that the SPM method for the anisotropic 2PCF can be evaluated using FTs (SE15), and Bianchi et al. (2015) and Scoccimarro (2015) show the same for the anisotropic power spectrum.  In the present work, we show how bisector and midpoint methods can also be evaluated with FTs.  We do so by using Taylor series to compute the difference between these methods and the SPM method, showing that at leading order the difference is $\oO(\theta^2)$. The bisector and midpoint methods also differ from each other at this order. We present a slight generalization of the SPM estimator that is still computable with FTs and can be translated to the bisector and midpoint methods including terms of $\oO(\theta^2)$. 

The present work is complementary to recent work by Samushia, Branchini \& Percival (2015) empirically comparing the flat-sky and SPM methods to the midpoint method.  For flat-sky they find a scale-independent multiplicative renormalization of the $l>0$ moments of the anisotropic power spectrum, which should translate to the same effect in the anisotropic 2PCF. For SPM they find a more complicated, scale-dependent difference from the midpoint method. They give correction formulae for both cases assuming a thin-shell spherical cap survey, that $\theta$ is small, and that boundary effects are negligible. Providing these assumptions hold, Samushia, Branchini \& Percival's formulae can be used to translate SPM to midpoint. One element of our work here is also such a conversion, but with no restrictions on survey geometry, triangle opening angle, or survey boundaries.

We close by calculating PT predictions for this generalized estimator including $\oO(\theta^2)$ wide-angle contributions from the full three-parameter geometry, from which the predictions for bisector and midpoint method can be straightforwardly found. These predictions improve upon the Kaiser/Hamilton formulae in two ways: they use more accurate lines of sight, and they include wide-angle, $\oO(\theta^2)$ terms.  While computed using a straightforward inverse FT of formulae in P\'apai \& Szapudi (2008), our predictions may be of more practical utility because they are presented in a basis in which the data can be measured using FTs.  Further, P\'apai \& Szapudi include a number of ``non-perturbative'' terms involving reciprocal trigonemetric functions of the triangle's angles; our work perturbatively orders these corrections in powers of $\theta$. 

Throughout this work, we present explicit expressions including only the leading $\oO(\theta^2)$ wide-angle corrections. However the techniques we develop, both for relating SPM and bisector/midpoint methods and for computing the PT predictions, are fully general and could be used to arbitrary order in $\theta$.  

While wide-angle corrections are not important at the Baryon Acoustic Oscillation scale of $100\Mpch$ in current surveys such as BOSS, working on larger scales such as those relevant for measurements of primordial non-Gaussianity (Dalal et al. 2008) will likely require these corrections, especially given the small amplitude of the signal being sought.

\section{Three possible lines of sight}
\label{sec:definitions}

\subsection{Single-Pair-Member (SPM) estimator}
\label{subsec:singlepair_def}
The SPM estimator computes the value of the 2PCF weighted by
Legendre polynomials of the angle between the line of sight to one
galaxy, given by the vector to that galaxy $\vec{r}_{i}$, and the
separation vector $\vec{s}$. We have

\begin{equation}
\xi_{l}^{\rm SP}(S)=\frac{1}{2}\sum_{i=1,2}\int d^{3}\vec{s} \;d^{3}\vec{r}_{i}\; \Phi(|\vs|;S) P_{l}(\hat{r}_{i}\cdot\hat{s})N(\vec{r}_{i})N(\vec{r}_{i}+\vec{s}).
\label{eqn:singlepair_def}
\end{equation}
$\Phi(|\vs|;S)$ bins separations into bins denoted by $S$. The sum above occurs because each pair of galaxies enters twice, once with sightline as $\vr_1$ and once with sightline as $\vr_2$. Thus the SPM estimator averages over lines
of sight after first weighting by the multipoles. As shown in SE15, decomposing the Legendre polynomial $P_{l}$
into spherical harmonics using the spherical harmonic addition theorem
allows evaluation of this estimator with FTs; Bianchi et al.
(2015) and Scoccimarro (2015) show the analogous result for the anisotropic power spectrum.

\subsection{Bisector estimator}
\label{subsec:bisec_def}
The bisector estimator computes the 2PCF weighted by Legendre polynomials in the cosine $\muB$ of the angle between the angle bisector of $\theta$ and the separation vector $\vs$. $\theta$ is the opening angle of the observer-galaxy-pair triangle; the geometry is shown in the lefthand panel of Figure \ref{fig:midpt_bisec_fig}. We have

\begin{align}
&\xi_{l}^{\rm B}(S)=\nonumber\\
&\int d^{3}\vec{s} \; d^{3}\vec{r}_{i}\; \Phi(|\vs|;S) P_{l}\left(\frac{1}{2}\left[\hat{r}_{1}+\hat{r}_2\right]\cdot\hat{s}\right)N(\vec{r}_{i})N(\vec{r}_{i}+\vec{s}).
\label{eqn:bisec_def}
\end{align}
Where the SPM estimator averages over lines of sight after projecting onto $P_l$, the bisector method does so before projecting. The difference between the two methods is simply produced by the extent to which these two operations fail to commute.

\subsection{Midpoint estimator}
\label{subsec:midpt_def}
The midpoint estimator computes the 2PCF weighted by Legendre polynomials in the cosine $\muM$ of the angle between the vector pointing from the observer to separation's midpoint and the separation. This geometry is shown in the righthand panel of Figure \ref{fig:midpt_bisec_fig}. We have
\begin{equation}
\xi_{l}^{\rm B}(S)=\int d^{3}\vec{s}\;d^{3}\vec{r}_{i}\;\Phi(|\vs|;S) P_{l}\left(\frac{\vec{r}_1+\vec{r}_2}{|\vec{r}_1+\vec{r}_2|}    \cdot\hat{s}\right)N(\vec{r}_{i})N(\vec{r}_{i}+\vec{s}).
\label{eqn:midpt_def}
\end{equation}
Where the bisector method took unit vectors to each galaxy and then averaged them, this method averages first and then takes the unit vector.  The difference between bisector and midpoint is thus again produced by a commutation failure.

\section{Relating the SPM, midpoint, and bisector estimators}
\label{sec:midpt_bisector_reln}
Here we show that the three methods of \S\ref{sec:definitions} only disagree at $\oO(\theta^2)$ and compute this difference explicitly. Figure 1 shows our notation: $d=s/2$ is the segment of $s$ cut by the vector $\vz$; the bisector cuts $s$ into two possibly unequal segments $d_1$ and $d_2$.  For the midpoint, $\epsilon \equiv d/z\ll 1$ for small $\theta$; indeed in this limit $\epsilon \approx \theta$.   $\epsilon_i \equiv d_i/z$ are the analogous parameters for the bisector, where $i$ ranges over $1, 2$.

\begin{figure}
\centering
\includegraphics[scale=0.3]{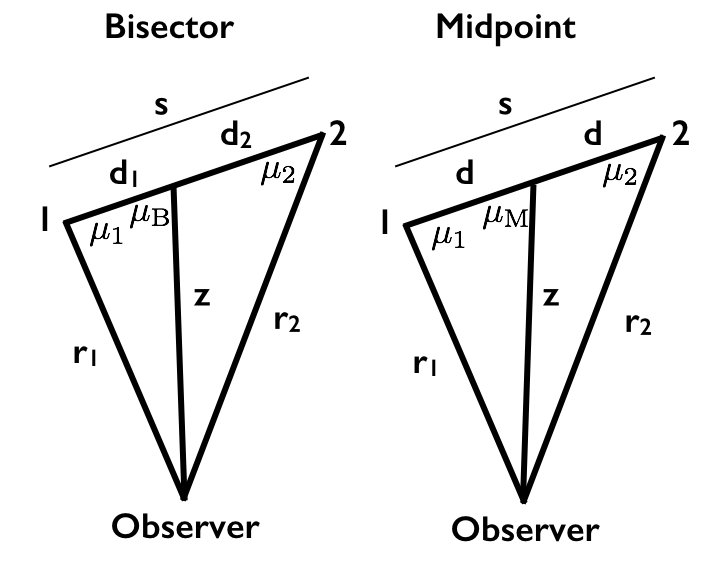}

\caption{Points $1$ and $2$ denote galaxies; the separation $\vs$ is the vector connecting them such that $\vr_1+\vs = \vr_2$. The lefthand panel shows the relevant parameters when the angle bisector at the observer's vertex is taken to be the line of sight $\vz$; the righthand panel the same when the vector from the observer to the separation's midpoint is used. $\muB$ and $\muM$ are $\hs\cdot\hz$ for the two estimators.}
\label{fig:midpt_bisec_fig}
\end{figure}

The SPM method measures $\frac{1}{2}\left[P_{l}(-\mu_{1})+P_{l}(\mu_{2})\right]$. We  begin by expanding this quantity in terms of both the midpoint parameters and the bisector parameters.  We present intermediate steps only for the midpoint geometry, but the results below will describe the bisector geometry as well with the replacements $d\to d_i$ and $\eps\to \eps_i$. Solving the law of cosines for $\mu_i$, we see from the lefthand panel of Figure \ref{fig:midpt_bisec_fig} that for both smaller triangles
\begin{align}
\mu_i =\frac{d^2+r_i^2-z^2}{2dr_i}
\label{eqn:mui}
\end{align}
where $i=1,2$. From the law of cosines
\begin{align}
r_i = z\left[ 1+ \eps^2\mp 2\eps \muM\right]^{1/2}
\end{align}
with minus sign for $i=1$ and plus for $i=2$.  Inserting this into equation (\ref{eqn:mui}) and simplifying we find
\begin{align}
\mu_i = \frac{\eps \mp \muM }{\sqrt{1+\eps^2\mp2\eps\muM}}
\label{eqn:mui_ito_eps}
\end{align}
Taking Taylor series for $\mu_i$ about $\muM$ yields
\begin{align}
\mu_{i} &=\mp\muM+\left(1-\muM^{2}\right)\epsilon\mp\frac{3}{2}\left(-\muM+\muM^{3}\right)\epsilon^{2}\nonumber\\
& +\left(-\frac{1}{2}+3\muM^{2}-\frac{5\muM^{4}}{2}\right)\epsilon^{3} + \oO(\eps^4).
\label{eqn:mui_series}
\end{align}

We now Taylor expand $P_l(\mp\mu_i)$ about $\muM$ and use equation (\ref{eqn:mui_series}) to then replace all $\mu_i$ in this series in terms of $\muM$, finding

\begin{align}
&P_{l}(\mp\mu_{i})\approx P_{l}(\muM)\nonumber\\
&+P_l'(\muM)\bigg[\mp\left(1-\muM^{2}\right)\epsilon+\frac{3}{2}\left(-\muM+\muM^{3}\right)\epsilon^{2}\bigg]\nonumber\\
&+\frac{1}{2}P_l''(\muM) \left(1-2\muM^{2}+\muM^{4}\right)\epsilon^{2}\mp \oO(\eps^3).
\end{align}
Forming the sum $\left[ P_l(-\mu_1)+P_l(\mu_2)\right]/2$ and using expressions derived along the same lines as above to write this sum also in terms of $\muB$ and $\eps_i$, we find
\begin{align}
&\frac{1}{2}\left[P_{l}(-\mu_{1})+P_{l}(\mu_{2})\right]=P_{l}(\mu_{M})+\bigg[ \frac{3}{2} P_l'(\muM)\left(-\mu_{M}+\mu_{M}^{3}\right) \nonumber\\
&+\frac{1}{2}P_l''(\muM)(1-2\mu_{M}^{2}+\mu_{M}^{4})\bigg]\epsilon^{2}+\mathcal{O}(\epsilon^{4})\nonumber\\
&= P_{l}(\mu_{B})+P_l'(\muB)\bigg[(1-\mu_{B}^{2})\frac{\epsilon_{2}-\epsilon_{1}}{2}+\frac{3}{2}\left(-\mu_{B}+\mu_{B}^{3}\right)\frac{\epsilon_{1}^{2}+\epsilon_{2}^{2}}{2}\bigg]\nonumber\\
&+\frac{1}{2}P_l''(\muB)\left[1-2\mu_{B}^{2}+\mu_{B}^{4}\right]\frac{\epsilon_{1}^{2}+\epsilon_{2}^{2}}{2}+\mathcal{O}(\epsilon_{i}^{4}).
\label{eqn:general_exp}
\end{align}
We have consolidated terms in $\eps_1^3-\eps_2^3$ into the $\oO(\eps_i^4)$ term, since differences of powers of small quantities always involve one power higher. Equation (\ref{eqn:general_exp}) already shows that midpoint, bisector, and SPM estimators agree at $\oO(\theta)$.  Notice also that in the limit where $\eps_1 = \eps_2$, the triangle in Figure 1 is isosceles and so midpoint and bisector are the same vector; $\muB = \muM$ and the above expression for the bisector simplifies to that for the midpoint. Also note that for the midpoint only even powers of $\eps$ enter its difference from the SPM method; this is due to parity.

Our goal is now to reduce equation (\ref{eqn:general_exp}) to a simple expression showing how $P_l(\muM)$ and $P_l(\muB)$ differ, so we must replace $\muB$ and $\eps_i$ with expressions in $\muM$ and $\eps$. We first relate $\eps_i$ to $\eps$: using their definitions it can be shown that
\begin{align}
\frac{\epsilon_1+\epsilon_2}{2}=\epsilon.
\label{eqn:epsi_eps}
\end{align}

We now obtain $\eps_2$ in terms of $\eps_1$ and then replace $\eps_1$ with $\eps$. The Angle Bisector Theorem means $d_1/r_1 = d_2/r_2$ for the lefthand triangle in Figure \ref{fig:midpt_bisec_fig} and replacing the $r_i$ using the law of cosines yields
\begin{equation}
\frac{d_{1}}{z\left[\epsilon_{1}^{2}+1-2\epsilon_{1}\mu_{B}\right]^{1/2}}=\frac{d_{2}}{z\left[\epsilon_{2}^{2}+1+2\epsilon_{2}\mu_{B}\right]^{1/2}}.
\label{eqn:bisector_step1}
\end{equation}
Notice the relative sign between the two sides for the cross term in $2\muB\eps_i$ in the denominator; this is because the two angles between $\vs$ and $\vz$ in the lefthand panel are supplementary.
Taylor expanding each side of equation (\ref{eqn:bisector_step1}) yields
\begin{align}
&\epsilon_{1}+\mu_{B}\epsilon_{1}^{2} \approx\epsilon_{2}-\mu_{B}\epsilon_{2}^{2}
\label{eqn:eps1_eps2}
\end{align}
Inserting these results in equation (\ref{eqn:general_exp}), retaining only $\oO(\eps^2)$ terms, and simplifying yields
\begin{align}
&P_{l}(\mu_{M}) + \nonumber\\
&\bigg[P_l'(\muM)\frac{3}{2}\left(-\mu_{M}+\mu_{M}^{3}\right)+\frac{1}{2}P_l''(\muM)\left(1-2\mu_{M}^{2}+\mu_{M}^{4}\right)\bigg]\epsilon^{2}\nonumber\\
&\approx P_{l}(\mu_{B}) +\bigg[ \frac{1}{2} P_l'(\muB)\left(-\mu_{B}+\mu_{B}^{3}\right)\nonumber\\
&+\frac{1}{2}P_l''(\muB)\left(1-2\mu_{B}^{2}+\mu_{B}^{4}\right)\bigg]\epsilon^{2}.
\label{eqn:midpt_bisec_reln_simp}
\end{align}

We now relate $\muB$ to $\muM$. Setting $l=1$ in equation (\ref{eqn:midpt_bisec_reln_simp}) gives
\begin{equation}
\mu_{M}+\frac{3}{2}\left(-\mu_{M}+\mu_{M}^{3}\right)\epsilon^{2}\approx\mu_{B}+\frac{1}{2}\left(-\mu_{B}+\mu_{B}^{3}\right)\epsilon^{2};
\label{eqn:muM_muB_trans}
\end{equation}
i.e., $\muM = \muB +\oO(\eps^2)$. Using this result in equation (\ref{eqn:midpt_bisec_reln_simp}) and rearranging shows that at leading order (which is $\eps^2$) the midpoint and bisector methods differ as
\begin{align}
P_l(\muM)-P_l(\muB) &\approx - P_l'(\muB)\left[ -\muB+\muB^3 \right]\eps^2\nonumber\\
& \approx - P_l'(\muM)\left[ -\muM+\muM^3 \right]\eps^2.
\label{eqn:bisec_midpt_diff}
\end{align}

\section{Midpoint and bisector methods via generalized SPM}
We now show that a generalization of the SPM estimator can be used to find the midpoint and bisector estimators to arbitrary accuracy in $\theta$. We then present explicit formulae including wide-angle corrections in $\theta^2$.  The generalized SPM estimator we present can be computed using FTs, meaning that both midpoint and bisector methods can be as well. 
Recalling equation (\ref{eqn:general_exp}) we already know that both midpoint and bisector methods at $\oO(\theta)$ {\it are} simply the SPM method; we now discuss how to cancel off the difference at $\oO(\theta^2)$. We first focus on the midpoint estimator and restrict attention to $l=2$. 

Writing out the term in $\eps^2$ on the midpoint side of equation (\ref{eqn:midpt_bisec_reln_simp})
explicitly, we have
\begin{align}
&\left[\frac{3}{2} P_2'(\muM)\left(-\muM+\muM^{3}\right)+\frac{1}{2}P_2''(\muM)\left(1-2\muM^{2}+\muM^{4}\right)\right]\epsilon^{2}\nonumber\\
&=\frac{1}{35}\left[7P_{0}\left(\muM\right)-55P_{2}\left(\muM\right)+48P_{4}(\muM)\right]\epsilon^{2}.
\label{eqn:midpt_error_term}
\end{align}
Each of the $P_l$ on the right hand side may be estimated with
error at $\oO(\eps^2)$ using equation (\ref{eqn:general_exp}) with $l=0,2,$ and $4$, and we recall equations (\ref{eqn:eps1_eps2}) and (\ref{eqn:epsi_eps}) to estimate $\eps^2$. 
Subtracting equation (\ref{eqn:midpt_error_term}) from equation (\ref{eqn:general_exp}) will then give the midpoint estimator accurately including $\oO(\eps^2)$ terms.

A key point is that in equation (\ref{eqn:midpt_error_term}), we required quantities of the form $P_l(\muM)\eps^n$. To $\oO(\eps^n)$, $\eps^n \approx \eps_i^n\approx s/(2r_i)$ from the geometry in Figure \ref{fig:midpt_bisec_fig}. This motivates defining the generalized SPM estimator
\begin{align}
&\xi_{ln}^{\rm SPG}(S)=\frac{1}{2}\sum_{i=1,2}\int d^{3}\vec{s} \;d^{3}\vec{r}_{i}\;\nonumber\\
&\times\Phi(|\vs|;S)\left(\frac{s}{r_i}\right)^nP_{l}(\hat{r}_{i}\cdot\hat{s})N(\vec{r}_{i})N(\vec{r}_{i}+\vec{s}).
\label{eqn:genest_noselxn_def}
\end{align}
This is just the SPM estimator weighted by a power of $s/r_i$ prior to integration. SE15 showed the $n=0$ case can be done with FTs, and $n>0$ does not change this conclusion. Explicitly we have
\begin{align}
&\xi_{ln}^{\rm SPG}(S)=\frac{4\pi}{2(2l+1)}\sum_{m=-l}^l\sum_{i=1,2}\int d^{3}\vec{s}\; \Phi(|\vs|;S) s^n Y_{lm}^*(\hs)\nonumber\\
&\times \int d^3\vr_i\; r_i^{-n} Y_{lm}(\hr_i)N(\vr_i)N(\vr_i+\vs)
\label{eqn:genest_convol}
\end{align}
by applying the spherical harmonic addition theorem, where the $Y_{lm}$ are spherical harmonics and star means conjugate. The integral over $d^3\vr_i$ is a convolution and thus can be computed with an FT.

In terms of $\xi_{ln}^{\rm SPG}$, the midpoint estimator at $l=2$ is
\begin{align}
&\xi_2^{\rm M}(S) =\nonumber\\
&\xi_{20}^{\rm SPG}(S) -\frac{1}{35}\bigg[ 7\xi_{02}^{\rm SPG}(S) -55 \xi_{22}^{\rm SPG}(S) + 48 \xi_{42}^{\rm SPG}(S)\bigg].
\end{align}

Using the same approach, one can find the midpoint for $l=4$ and $l=6$:
\begin{align}
\xi_4^{\rm M}(S) &=\xi_{40}^{\rm SPG}(S)\nonumber\\
& -\frac{10}{77}\bigg[ 11 \xi_{22}^{\rm SPG}(S) -39 \xi_{42}^{\rm SPG}(S)  + 28 \xi_{62}^{\rm SPG}(S) \bigg];\nonumber\\
\xi_6^{\rm M}(S) &=\xi_{60}^{\rm SPG}(S)\nonumber\\
& -\frac{7}{715}\bigg[ 375 \xi_{42}^{\rm SPG}(S) -1079 \xi_{62}^{\rm SPG}(S)  + 704 \xi_{82}^{\rm SPG}(S) \bigg].
\end{align}
We see that removing the $\mathcal{O}(\epsilon^{2})$ errors has a price:
we need to compute more multipoles, in particular those at $l\pm2$. 

Again using equation (\ref{eqn:midpt_bisec_reln_simp}) one can find the bisector including terms in $\theta^2$ as
\begin{align}
\xi_2^{\rm B}(S) &=\xi_{20}^{\rm SPG}(S)\nonumber\\
& -\frac{1}{35}\bigg[ 21 \xi_{02}^{\rm SPG}(S) -45 \xi_{22}^{\rm SPG}(S)  + 24 \xi_{42}^{\rm SPG}(S) \bigg];\nonumber\\
\xi_4^{\rm B}(S) &=\xi_{40}^{\rm SPG}(S)\nonumber\\
& -\frac{1}{231}\bigg[ 550 \xi_{22}^{\rm SPG}(S) -1110 \xi_{42}^{\rm SPG}(S)  + 560 \xi_{62}^{\rm SPG}(S) \bigg];\nonumber\\
\xi_6^{\rm M}(S) &=\xi_{60}^{\rm SPG}(S)\nonumber\\
& -\frac{1}{715}\bigg[ 3675 \xi_{42}^{\rm SPG}(S) -7371 \xi_{62}^{\rm SPG}(S)  + 3696 \xi_{82}^{\rm SPG}(S) \bigg].
\end{align}

\section{Perturbation theory predictions}
\label{sec:PT_predictions}

We now wish to compute the expectation value of the generalized estimator (\ref{eqn:genest_noselxn_def}) from
linear perturbation theory. Incorporating a radial selection function $\phi$ we have

\begin{align}
&\left< \zeta^{\rm SPG}_ {ln}(S)\right>=\frac{1}{2}\sum_{\alpha=i,j}\int r_{\alpha}^{2}dr_{\alpha}\;\phi(r_{\alpha})\nonumber\\
&\times \int d^{3}\vec{s}\;\Phi(|\vs|;S)\left(\frac{s}{r_{\alpha}}\right)^{n}P_{l}(\hat{r}_{\alpha}\cdot\hat{s})\left<\delta_{{\rm s}}(\vec{r}_{i})\delta_{{\rm s}}\left(\vec{r}_{j}\right)\right>.
\label{eqn:SPG_wslxn}
\end{align}
From P\'apai and Szapudi (2008) the expectation value density field's 2PCF is
\begin{equation}
\left<\delta_{{\rm s}}(\vec{r}_{i})\delta_{{\rm s}}(\vec{r}_{j})\right>=\int\frac{d^{3}\vec{k}}{\left(2\pi\right)^{3}}P(k)e^{-i\vec{k}\cdot\vec{s}}\kappa(k,r_{i},r_{j},\hat{x}_{i}\cdot\hat{k},\hat{x}_{j}\cdot\hat{k})
\end{equation}
where $P(k)$ is the linear power spectrum, $\delta_{\rm s}$ is the redshift space density perturbation, and
\begin{align}
&\kappa(k,r_{i},r_{j},\hat{r}_{i}\cdot\hat{k},\hat{r}_{j}\cdot\hat{k})\equiv \nonumber\\
& \bigg\{\left(1+\frac{f}{3}\right)^{2}+2f\left(1+\frac{f}{3}\right)\bigg[\frac{1}{3}P_{2}(\hat{r}_{i}\cdot\hat{k})+\frac{1}{3}P_{2}(\hat{r}_{j}\cdot\hat{k})\nonumber\\
&+\frac{i}{kr_{i}}P_{1}(\hat{r}_{i}\cdot\hat{k})-\frac{i}{kr_{j}}P_{1}(\hat{r}_{j}\cdot\hat{k})\bigg]\nonumber\\
&+4f^{2}\bigg[\frac{1}{9}P_{2}(\hat{r}_{i}\cdot\hat{k})P_{2}(\hat{r}_{j}\cdot\hat{k})+\frac{1}{k^{2}r_{i}r_{j}}P_{1}(\hat{r}_{i}\cdot\hat{k})P_{1}(\hat{r}_{j}\cdot\hat{k})\nonumber\\
&+\frac{i}{kr_{i}}P_{1}(\hat{r}_{i}\cdot\hat{k})P_{2}(\hat{r}_{j}\cdot\hat{k})-\frac{i}{kr_{j}}P_{1}(\hat{r}_{j}\cdot\hat{k})P_{2}(\hat{r}_{i}\cdot\hat{k})\bigg]\bigg\}.
\end{align}
$f=d\ln D/d\ln a \propto \omegam^{0.6}$ is the logarithmic derivative of the linear growth rate $D$ with respect to scale factor $a$.
Without loss of generality we consider $\alpha=i.$ Our goal now is to simplify this integral into one-dimensional Hankel transforms of the power spectrum.  First, notice that all terms involving only $\vr_i$ can be evaluated by applying the plane wave expansion (Arfken, Weber, \& Harris 2013, hereafter AWH13, equation 16.63) and the spherical harmonic addition theorem (NIST DLMF 14.30.9) to write all angular dependence in terms of spherical harmonics. We denote these terms with a subscript ``$\vr_i$ only.'' Performing the angular integral and invoking orthogonality yields 
\begin{align}
&\left<\delta_{{\rm s}}(\vec{r}_{i})\delta_{{\rm s}}(\vec{r}_{j})\right>_{\vr_i\;{\rm only}} = \left(1+\frac{f}{3}\right)^2\xi_0(S)\nonumber\\
& +2f\left(1+\frac{f}{3}\right)\bigg[-\frac{1}{3}\xi_2(s)P_2(\mu_i)+\frac{1}{r_i}\xi_1^{[-1]}(s)P_1(\mu_i)\bigg] 
\label{eqn:ri_only}
\end{align}
where 
\begin{align}
&\xi_l (s) \equiv \int \frac{k^2dk}{2\pi^2}P(k)j_l(kr_i),\nonumber\\
&\xi^{[t]}_l (s) \equiv \int \frac{k^2dk}{2\pi^2}k^{t}P(k)j_l(kr_i)
\end{align}
and $\mu_i \equiv \hr_i\cdot\hs$.

More difficult to evaluate are the terms involving $\hat{r}_{j}=(\vec{r}_{i}+\vec{s})/|\vec{r}_{i}+\vec{s}|$ or $1/r_j$; we will denote these terms with a subscript ``$\vr_j$ enters.'' We can expand these terms perturbatively in powers
of $\epsilon_{i}\equiv s/r_{i}$ and $\hat{r}_{i}\cdot\hat{s}$ and then perform the integral over $\vec{k}$. Here and in what follows we work only to $\oO(\eps_i^2)\sim\theta^2$, though the techniques here could be used to go to higher order if desired. We find
\begin{align}
\frac{1}{r_j} &=\frac{1}{\left[1+\eps_i^2+2\mu_i\eps_i \right]^{1/2}}\nonumber\\
&\approx 1-\mu_i \eps_i +\left(-\frac{1}{2}+\frac{3}{2}\mu_i^2 \right)\eps_i^2
\end{align}
and from this
\begin{align}
\hr_j &= \frac{\hr_i + \eps_i \hs}{r_j}\nonumber\\
& \approx \hr_i \left[1-\mu_i \eps_i +\left(-\frac{1}{2}+\frac{3}{2}\mu_i^2 \right)\eps_i^2 \right]+\hs\left[\eps_i-\mu_i\eps_i^2 \right].
\label{eqn:rj_hat}
\end{align}
We now need the $P_l(\hr_j\cdot\hk)$.  These can be computed explicitly by dotting equation (\ref{eqn:rj_hat}) with $\hk$ and evaluating the required powers; consequently they will depend on $\eps_i$, $\mu_i$, $\hr_i\cdot\hk \equiv \tilde{\mu}_i$, and $\hs\cdot\hk \equiv \tilde{\mu}$.  To track powers of our perturbative quantity $\eps_i$ and retain compact notation, we write the $P_l(\hr_j\cdot\hk)$ as the triple Legendre series
\begin{align}
& P_{l}(\hat{x}_{j}\cdot\hat{k})=\sum_{abcd}\mathcal{P}_{abcd}^{[l]}\epsilon_{i}^{a}P_{b}(\mu_{i})P_{c}(\tilde{\mu}_{i})P_{d}(\tilde{\mu})
\label{eqn:legendre_series_for_p}
\end{align}
with constant coefficients $\mathcal{P}^{[l]}_{abcd}$ given in Table \ref{table:pcals}. We emphasize that the first index refers to the power of the small parameter $\eps_i$ entering our expansion, so for large $a$ the terms in the series become negligible. 

\begin{table}
\begin{tabular}{|c|c|c|}
\hline 
$\mathcal{P}_{0010}^{[1]}=1$ & $\mathcal{P}_{1110}^{[1]}=-1$ & $\mathcal{P}_{1001}^{[1]}=1$\tabularnewline
$\mathcal{P}_{2210}^{[1]}=1$ & $\mathcal{P}_{2101}^{[1]}=-1$ &\tabularnewline
\hline 
\end{tabular}
\\
\\
\\
\begin{tabular}{|c|c|c|c|}
\hline 
$\mathcal{P}_{0000}^{[2]}=-1$ & $\mathcal{P}_{0020}^{[2]}=1$ & $\mathcal{P}_{1011}^{[2]}=3$ & $\mathcal{P}_{1100}^{[2]}=-1$\tabularnewline
$\mathcal{P}_{1120}^{[2]}=-2$ & $\mathcal{P}_{2002}^{[2]}=1$ & $\mathcal{P}_{2111}^{[2]}=-3$& $\mathcal{P}_{2020}^{[2]}=-1$\tabularnewline
$\mathcal{P}_{2220}^{[2]}=4$ & $\mathcal{P}_{2000}^{[2]}=1$ & $\mathcal{P}_{2021}^{[2]}=-2$ & \tabularnewline
\hline 
\end{tabular}
\caption{Coefficients in the Legendre series for $P_l(\hr_j\cdot\hk)$ equation (\ref{eqn:legendre_series_for_p}).}
\label{table:pcals}
\end{table}

We now adopt the same form of series for $\kappa$ but add an additional index to track powers of $k$, and promote the coefficients in the series to depend on $r_i$, finding
\begin{align}
&\kappa_{\vx_j\;{\rm enters}}(k,x_{i},x_{j},\hat{x}_{i}\cdot\hat{k},\hat{x}_{j}\cdot\hat{k})=\nonumber\\
&\sum_{rtuvw}\kappa_{rtuvw}(x_{i})\epsilon_{i}^{r}k^{-t}P_{u}(\mu_{i})P_{v}(\tilde{\mu}_{i})P_{w}(\tilde{\mu}).
\label{eqn:kappa_off_series}
\end{align}
The coefficients of terms with no $k$ dependence are
\begin{align}
&\kappa_{a0bld}=\frac{2f}{3}\left(1+\frac{f}{3}\right)\mathcal{P}_{abcd}^{[2]}\delta_{cl'}^{K}\nonumber\\
&+\frac{4f^{2}}{9}(2l+1)\sum_{c}\left(\begin{array}{ccc}
2 & c & l\\
0 & 0 & 0
\end{array}\right)^{2}\mathcal{P}_{abcd}^{[2]};
\end{align}
recall that $c\leq 2$ meaning $l\leq 4$. In this computation we linearized the product of two Legendre polynomials into a sum over one generating a Wigner 3j-symbol.  Note that the coefficients above could have also been obtained by direct computation of polynomials in $\tilde{\mu}$, $\tilde{\mu}_i$, and $\mu$. However, when expanded terms such as $P_2(\tilde{\mu}_i)P_2((\tilde{\mu}_i)$  produce a large number of terms, so we have found the method presented above preferable in practice. 

In contrast, for the terms involving $1/k$ and $1/k^2$, we found it simpler to use direct computation because these terms have lower-$l$ Legendre polynomials and so involve fewer terms.  Consequently we did not obtain a general formula for these coefficients. We simply present them in Tables \ref{table:one_over_k_terms} and \ref{table:one_over_ksq_terms} above.
\begin{table}
\begin{tabular}{|l|}
\hline
$\kappa_{\vr_j\;{\rm enters}}$ coefficients $\propto 1/k$\tabularnewline
\hline 
$\kappa_{01010}=i[2f(1+f/3)+2f^2]/r_i$\tabularnewline
$\kappa_{11001}=i[2f(1+f/3)+4f^2]/r_i$\tabularnewline
$\kappa_{11110}=i[4f(1+f/3)-4f^2]/r_i$\tabularnewline 
$\kappa_{11021}=4if^2/r_i$\tabularnewline
$\kappa_{21101}=i[4f(1+f/3)-8f^2]/r_i$\tabularnewline
$\kappa_{21010}=i[2f(1+f/3)+2f^2]/r_i$\tabularnewline
$\kappa_{21210}=i[16f(1+f/3)/3-16f^2/3]/r_i$\tabularnewline
$\kappa_{21010}=i[8f(1+f/3)/3-8f^2/3]/r_i$\tabularnewline
$\kappa_{21010}=4if^2/r_i$\tabularnewline 
$\kappa_{21121}=-8if^2/r_i$\tabularnewline
\hline
\end{tabular}
\caption{Coefficients for $1/k$ terms in equation (\ref{eqn:kappa_off_series}).}
\label{table:one_over_k_terms}
\end{table}

\begin{table}
\begin{tabular}{|l|}
\hline
$\kappa_{\vr_j\;{\rm enters}}$ coefficients $\propto 1/k^2$\tabularnewline
\hline 
$\kappa_{02020}=8f^2/(3r_i^2)$\tabularnewline
$\kappa_{02000}=4f^2/(3r_i^2)$\tabularnewline
$\kappa_{12011}=4f^2/r_i^2$\tabularnewline
$\kappa_{12120}=-16f^2/(3r_i^2)$\tabularnewline
$\kappa_{12100}=-8f^2/(3r_i^2)$\tabularnewline
$\kappa_{22111}=-8f^2/r_i^2$\tabularnewline
\hline
\end{tabular}
\caption{Coefficients for $1/k^2$ terms in equation (\ref{eqn:kappa_off_series}).}
\label{table:one_over_ksq_terms}
\end{table}

With $\kappa_{\vr_j\;{\rm enters}}$ written in the form (\ref{eqn:kappa_off_series}) we apply a theorem proven in the Appendix to find
\begin{align}
&\left<\delta_{{\rm s}}(\vec{r}_{i})\delta_{{\rm s}}(\vec{r}_{j})\right>_{\vr_j\;{\rm enters}}=\nonumber\\
&\sum_{rtuvw}\kappa_{rtuvw}(r_{i})\epsilon_{i}^{r}P_{u}(\mu_{i})\nonumber\\
&\times P_{v}(\mu_{i})\sum_{l'}(-i)^{l'}(2l'+1)^2\left(\begin{array}{ccc}
v & w & l'\\
0 & 0 & 0
\end{array}\right)^{2}\xi_{l'}^{[-t]}(s)
\end{align}
which can be further simplified by linearizing the Legendre polynomials to yield
\begin{align}
&\left<\delta_{{\rm s}}(\vec{r}_{i})\delta_{{\rm s}}(\vec{r}_{j})\right>_{\vr_j\;{\rm enters}}=\nonumber\\
&\sum_{L}\sum_{rtuvw}\kappa_{rtuvw}(r_{i})\epsilon_{i}^{r}\nonumber\\
&\times \sum_{l'}(-i)^{l'}(2l'+1)^2\left(\begin{array}{ccc}
v & w & l'\\
0 & 0 & 0
\end{array}\right)^{2}\xi_{l'}^{[-t]}(s)\nonumber\\
&\times (2L+1)\left(\begin{array}{ccc}
u & v & L\\
0 & 0 & 0
\end{array}\right)^{2}P_{L}(\mu_{i})
\label{eqn:rj_only}
\end{align}
Returning to equation (\ref{eqn:SPG_wslxn}) and performing the angular integral over $d\Omega_{s}$ against $P_{l}(\mu_{i})$ sets $L=l$ by orthogonality and gives a factor of $2/(2l+1)$. From the coefficients of $\kappa_{\vx_j\;{\rm enters}}$, working only to $\oO(\theta^2)$ we see that $u_{{\rm max}}=2,$
$v_{{\rm max}}=l'_{{\rm max}}=4,$ and $w_{{\rm max}}=d_{{\rm max}}=2.$
The angular momentum couplings in the 3j-symbols above may be illustrated
by triangle diagrams (see e.g. Brink \& Satchler 1993), from which
we see $L_{{\rm max}}=6$ and $l'_{{\rm max}}=6.$ 

Finally, one can combine equations (\ref{eqn:ri_only}) and (\ref{eqn:rj_only}) to give the full $\left<\delta_{\rm s}(\vr_i)\delta_{\rm s}(\vr_j)\right>$. Once we are given the binning $\Phi(|\vs|;S)$ and selection function $\phi(r_i)$,
we can carry out the integrals of equation (\ref{eqn:SPG_wslxn}); the integral over $r_i$ converts
the $r_{i}$ dependence of each term to a simple
numerical constant that encodes the survey geometry. This will give $\left<\xi^{\rm SPG}_{ln}(S)\right>$, which depends only on the separation bin $S$
and, more importantly, on $f$, the logarithmic derivative of the linear growth rate. Hence by measuring a number of $\xi_{ln}$ from the data for different $l$ and $n$
we obtain a system of equations involving the power spectrum
and $f$ that can be solved for both.

\section{Conclusions}
We have presented a framework that permits using FTs to measure the anisotropic 2PCF using either the angle bisector or the vector to the separation's midpoint as the line of sight to each galaxy pair.  We first showed that bisector, midpoint, and single pair member (SPM) methods only disagree beginning at $\oO(\theta^2)$ and then that a slight generalization of the SPM method could be used to remove this disagreement.  This generalized SPM method can be evaluated using FTs and so allows computation of the bisector or midpoint methods, in principle to arbitary order in $\theta$, using FTs.  Finally, we presented PT predictions for this generalized estimator including terms in $\theta^2$.  These predictions can be easily translated to predictions for the bisector and midpoint methods using our earlier work; this shows the error induced if wide-angle corrections are neglected.  

A number of previous works have predicted the wide-angle corrections to the anisotropic power spectrum or 2PCF (Fisher, Scharf \& Lahav 1994; Zaroubi \& Hoffman 1994; Heavens \& Taylor 1994; Tegmark \& Bromley 1995; Hamilton \& Culhane 1996; Szalay 1998; Bharadwaj 1999; Taylor \& Valentine 1999; Matsubara 2000; Szapudi 2004; P\'apai \& Szapudi 2008; Reimberg, Bernardeau \& Pitrou 2015); indeed P\'apai \& Szapudi (2008) was our starting point here. The key advance of our work is casting this prediction in terms of an estimator that can be easily measured using FTs, as well as ordering our prediction in a perturbative series in powers of $\theta$. This ordering shows the error induced when RSD are measured using only 2 parameters (separation and angle between the separation and the line of sight) rather than 3 (separation and two lines of sight or an equivalent combination).  Thus if the maximum opening angle of the triangles in a given survey were known, an upper bound could be simply placed on the error induced by ignoring wide-angle corrections.  Even more accurate would be to estimate this error by integrating $\theta^2$ against the probability distribtuion of triangles in the survey as a function of $\theta$.

In future work, we will implement the methods described here to produce an anisotropic 2PCF algorithm allowing computation of all standard line of sight definitions via FTs.  Using our PT prediction of \S\ref{sec:PT_predictions} it will then be straightforward to extract from data a measurement of $f\sigma_8$ that is more accurate than those obtained using the Kaiser/Hamilton formulae.  

Given the large number of objects and large volumes of upcoming surveys such as DESI and Euclid, it will be desirable to have FT methods for computing the anisotropic 2PCF that permit use of any desired definition of the line of sight. Since gridding the data does lose some spatial information, we expect that Fourier techniques will complement rather than replace pair counting in this context, being especially useful where the trade-off between spatial accuracy and a large number of objects or catalogs is favorable.   The speed of the approach presented here should be particularly helpful in using thousands of mock catalogs to compute covariance matrices, in doing edge correction which requires computing the anisotropic 2PCF for of order $\sim 100$ random catalogues per dataset, and in testing PT predictions against N-body simulations.

\section*{Acknowledgments}

ZS thanks Istv\'an Szapudi, Taka Matsubara, Lado Samushia, and Florian Beutler
for useful correspondence, and Will Percival, Davide Bianchi, Anthony
Challinor, Stephen Portillo, Charles-Antoine Collins-Fekete, and Alexander
Wiegand for useful conversations. ZS especially thanks Stephen Portillo for a careful read of the manuscript. This material is based upon work supported by the National Science Foundation Graduate Research Fellowship under Grant No. DGE-1144152; DJE is supported by grant ${\rm DE}$-SC0013718 from the U.S. Department of Energy.

\section*{References}

\hangindent=1.5em
\hangafter=1
\noindent Anderson L et al., 2014, MNRAS 441, 1, 4-62.

\hangindent=1.5em
\hangafter=1
\noindent Arfken GB, Weber HJ \& Harris FE, 2013, Mathematical Methods for Physicists: Academic Press, Waltham, MA.

\hangindent=1.5em
\hangafter=1
\noindent Beutler F et al., 2012, MNRAS, 423, 4, 3430-3444.

\hangindent=1.5em
\hangafter=1
\noindent Bharadwaj S, 1999, ApJ, 516, 2, 507-518.

\hangindent=1.5em
\hangafter=1
\noindent Bharadwaj S, 2001, MNRAS 327, 2, 577-587.

\hangindent=1.5em
\hangafter=1
\noindent Bianchi D, Gil-Mar\'in H, Ruggeri R \& Percival WJ, 2015, MNRAS 453, 1, L11-L15.

\hangindent=1.5em
\hangafter=1
\noindent Chuang C-H \& Wang Y, 2013, MNRAS 435, 1, 255-262.

\hangindent=1.5em
\hangafter=1
\noindent Dalal N, Dor\'e O, Huterer D \& Shirokov A, 2008, PRD 77, 12, 123514.

\hangindent=1.5em
\hangafter=1
\noindent Fisher KB, Scharf CA \& Lahav O, 1994, MNRAS, 266, 219.

\hangindent=1.5em
\hangafter=1
\noindent Hamilton AJS, 1992, ApJ, 385, L5.

\hangindent=1.5em
\hangafter=1
\noindent Hamilton AJS \& Culhane M, 1996, MNRAS 728, 73.

\hangindent=1.5em
\hangafter=1
\noindent Jackson JC, 1972, MNRAS, 156, 1.

\hangindent=1.5em
\hangafter=1
\noindent Kaiser N, 1987, MNRAS, 227, 1.

\hangindent=1.5em
\hangafter=1
\noindent Matsubara T, 2000, ApJ 535, 1, 1-23.

\hangindent=1.5em
\hangafter=1
\noindent Nishioka H \& Yamamoto K, 1999, ApJ 520:426-436.

\hangindent=1.5em
\hangafter=1
\noindent Oka A, Saito S, Nishimichi T, Taruya A \& Yamamoto K, 2014, MNRAS 439, 3, 2515-2530.

\hangindent=1.5em
\hangafter=1
\noindent P\'apai P \& Szapudi I, 2008, MNRAS 389, 1, 292-296.

\hangindent=1.5em
\hangafter=1
\noindent Peacock JA \& Dodds SJ, 1996, MNRAS, 280, L19.

\hangindent=1.5em
\hangafter=1
\noindent Peebles PJE, 1980, The Large-Scale Structure of the Universe: Princeton University Press, Princeton.

\hangindent=1.5em
\hangafter=1
\noindent Percival WJ \& White M, 2009, MNRAS 393, 1, 297-308.

\hangindent=1.5em
\hangafter=1
\noindent Raccanelli A, Samushia L \& Percival WJ, 2010, MNRAS 409, 4, 1525-1533.

\hangindent=1.5em
\hangafter=1
\noindent Reimberg P, Bernardeau F \& Pitrou C, 2015, preprint (arXiv:1506.06596).

\hangindent=1.5em
\hangafter=1
\noindent Brink DM \& Satchler GR, 1993, Angular Momentum: Oxford University Press, Oxford, UK.

\hangindent=1.5em
\hangafter=1
\noindent Samushia L, Percival WJ \& Raccanelli A, 2012, MNRAS 420, 3, 2102-2119.

\hangindent=1.5em
\hangafter=1
\noindent Samushia L, Branchini E \& Percival WJ, 2015, MNRAS, 452, 4, 3704-3709.

\hangindent=1.5em
\hangafter=1
\noindent Scoccimarro R, 2004, PRD, 70, 083007.

\hangindent=1.5em
\hangafter=1
\noindent Scoccimarro R, 2015, preprint (arXiv:1506.02729).

\hangindent=1.5em
\hangafter=1
\noindent Slepian Z \& Eisenstein DJ, 2015, MNRAS in press, arXiv:1506.04746. 

\hangindent=1.5em
\hangafter=1
\noindent Szalay A, Matsubara T \& Landy S, 1998, ApJ 498, 1, L1-L4.

\hangindent=1.5em
\hangafter=1
\noindent Szapudi I, 2004, ApJ, 614, 1, 51-55.

\hangindent=1.5em
\hangafter=1
\noindent Taddei L \& Amendola L, 2015, JCAP 2.

\hangindent=1.5em
\hangafter=1
\noindent Taruya A, Nishimichi T \& Saito S, 2010, PRD, 82, 063522.

\hangindent=1.5em
\hangafter=1
\noindent Taruya A, Koyama K, Hiramatsu T \& Oka A, 2014, PRD 89, 4, 043509.

\hangindent=1.5em
\hangafter=1
\noindent Taylor AN \& Hamilton AJS, 1996, MNRAS, 282, 3, 767-778.

\hangindent=1.5em
\hangafter=1
\noindent  Taylor AN \& Valentine H, 1999, MNRAS 306, 2, 491.

\hangindent=1.5em
\hangafter=1
\noindent Tegmark M \& Bromley B, 1995, ApJ 453, 533.%

\hangindent=1.5em
\hangafter=1
\noindent Yamamoto K, Nakamichi M, Kamino A, Bassett BA, Nishioka
H, 2006, PASJ, 58, 93.

\hangindent=1.5em
\hangafter=1
\noindent  Yoo J \& Seljak U, 2015, MNRAS, 447, 2, 1789-1805.

\hangindent=1.5em
\hangafter=1
\noindent Zaroubi S \& Hoffman Y, 1994,  preprint (arXiv:astro-ph/9311013).

\section*{Appendix}

Here we prove a theorem used to evaluate the $\kappa_{\vr_j\;{\rm enters}}$ terms of the PT predictions in \S\ref{sec:PT_predictions}, given by equation (\ref{eqn:kappa_off_series}). We show that 
\begin{align}
& I_{ujq}(s,\mu_{i})\equiv\int\frac{d^{3}\vec{k}}{\left(2\pi\right)^{3}}P(k)k^{-u}e^{-i\vec{k}\cdot\vec{s}}P_{j}(\tilde{\mu}_{i})P_{q}(\tilde{\mu})\nonumber\\
&=P_{j}(\mu_{i}) \sum_{l}(-i)^{l}(2l+1)^2\left(\begin{array}{ccc}
l & j & q\\
0 & 0 & 0
\end{array}\right)^{2}\xi_{l}^{[-u]}(s),
\label{eqn:theorem}
\end{align}
To show the theorem we
expand all Legendre polynomials in spherical harmonics via the spherical
harmonic addition theorem and expand the plane
wave into spherical harmonics and spherical Bessel functions using
AWH13 equation 16.63. Integration over $d\Omega_{k}$ then leaves
\begin{align}
& I_{ujq}(s,\mu_{i})=\frac{(4\pi)^2}{(2j+1)(2q+1)}\sum_{lmm'm''}(-i)^{l}\xi_{l}^{[-u]}(s)Y_{qm''}^{*}(\hat{s})\nonumber\\
&\times Y_{l m}^{*}(\hat{s})Y_{jm'}^{*} (\hat{r}_i)\mathcal{C}_{jql}\left(\begin{array}{ccc}
j & q & l\\
0 & 0 & 0
\end{array}\right)\left(\begin{array}{ccc}
j & q & l\\
m' & m'' & m
\end{array}\right)
\end{align}
with $\mathcal{C}_{jql}\equiv \sqrt{(2j+1)(2q+1)(2l+1)/(4\pi)}$. Inserting
NIST DLMF 34.3.20 to write the product $Y_{lm}^{*}(\hat{s})Y_{qm''}^{*}(\hat{s})$
as a sum over one spherical harmonic, summing the 3j-symbols over
$m$ and $m''$, and using the orthogonality identity NIST DLMF 34.3.16
we find
\begin{align}
I_{ujq}(s,\mu_{i})&=4\pi (2l+1)\sum_{l}(-i)^{l}\xi_{l}^{[-u]}(s)\left(\begin{array}{ccc}
j & q & l\\
0 & 0 & 0
\end{array}\right)^{2}\nonumber\\
&\times \sum_{m'=-j}^{j}Y_{jm'}^{*}(\hat{r}_{i})Y_{jm'}(\hat{s}).
\end{align}
Using the spherical harmonic addition theorem, the result (\ref{eqn:theorem}) follows immediately. 
\end{document}